\documentclass[aps,prb,10pt,twocolumn]{revtex4-2}

\usepackage{graphicx}
\usepackage{dcolumn}
\usepackage{color}
\usepackage{amsmath}
\usepackage{xfrac}
\usepackage{amssymb}
\usepackage{bm,dsfont}
\usepackage{multirow}
\usepackage{xcolor}
\usepackage[pdftex,colorlinks=true,linkcolor=blue,citecolor=magenta,filecolor=blue]{hyperref}
\usepackage{comment}

\newcommand{\ff}[1]{{\boldsymbol #1}}
\newcommand{\ca}[1]{{\cal #1}}
\newcommand{\bi}{\begin{itemize}}
\newcommand{\ei}{\end{itemize}}
\newcommand{\be}{\begin{equation}}
\newcommand{\ee}{\end{equation}}
\newcommand{\ba}{\begin{eqnarray}}
\newcommand{\ea}{\end{eqnarray}}

\newcommand{\refeq}[1]{Eq.\ (\ref{eq:#1})}
\newcommand{\labeq}[1]{\label{eq:#1}}

\renewcommand{\Im}{\operatorname{Im}}

\newcommand{\pp}[1]{{\color{blue}#1.}}

\newcommand{\titlepaper}{Turning Number and the Geometry of Disorder-Induced Topology}

\begin{document}

\title{\titlepaper}

\author{Lars Emmrich}
\author{Michael Potthoff}
\affiliation{Department of Physics, University of Hamburg, Notkestra{\ss}e 9-11, 22607 Hamburg, Germany}

\date{\today}

\begin{abstract}
The geometry of momentum-space Hamiltonian maps derived from the zero-frequency single-particle Green's function provides a route to a refined topological characterization of quantum systems beyond the noninteracting clean limit.
The central idea is to complement standard topological invariants by invariants that characterize the geometry of the map itself.
As a paradigmatic example, we study the one-dimensional chiral Su-Schrieffer-Heeger model with disorder, where the topological Hamiltonian defines a planar Bloch loop.
In addition to its conventional winding number, this loop possesses an integer turning number that characterizes the winding of its tangent vector.
We identify turning-number transitions that occur while the topological Hamiltonian remains finite and spectrally gapped, driven solely by a vanishing tangent vector of the Bloch loop, and obtain an extended phase diagram with additional topological structure beyond the winding-number classification.
Asymptotic analysis predicts the emergence and disorder dependence of the corresponding transition regime, in agreement with the numerical results.
Our findings establish the geometry of momentum-space Hamiltonian maps as an additional source of topological structure beyond conventional classifications.
\end{abstract}

\maketitle

\pp{Motivation}
The traditional classification of phases of matter is based on symmetry and spontaneous symmetry breaking, as embodied in Landau's paradigm \cite{Lan37a,Lan37b,Wen07}.
Topological phases extend this framework by allowing distinct phases that cannot be distinguished by local order parameters or symmetry breaking.

A major achievement in the theory of topological matter is the systematic classification of gapped noninteracting lattice-fermion systems \cite{SRFL08,Kit09,RSFL10}.
For each spatial dimension and each combination of time-reversal, particle-hole, and chiral symmetries, gapped Bloch Hamiltonians are assigned topological invariants that are unchanged under continuous deformations preserving both the bulk gap and the relevant symmetries.
Different values of these invariants therefore identify distinct topological phases.
Conversely, within the stable classification, Hamiltonians with the same invariants can be continuously deformed into one another without closing the gap or breaking the relevant symmetries.
Thus, within this framework, the classification is complete.

This completeness, however, refers to classification under continuous deformations and therefore does not resolve geometric properties of the Hamiltonian map that can change without changing its homotopy class. 
Can such geometric information itself carry robust topological structure? 
This question suggests a route distinct from existing refinements beyond the stable tenfold-way classification, including finite-band or fragile topology \cite{KG15,PWV18}, multigap topology \cite{JBL+21}, and crystalline classifications \cite{BEC+17}. 
These approaches retain additional band, gap, or symmetry information while leaving the underlying notion of continuous deformation essentially unchanged.

For interacting \cite{Rac18,Sen15} or disordered systems \cite{VP26}, topological phases can likewise be characterized by Green-function topological invariants, including the Ishikawa-Matsuyama invariant and its generalizations \cite{IM86,Vol03,Gur11,WZ12,WY13}. 
These are homotopy invariants of the single-particle Green’s function $\ff G(\ff k,i \omega)$ and remain unchanged under continuous deformations as long as neither poles nor zeros of $\ff G$ are encountered in frequency-momentum space.
Unlike in the noninteracting or clean case, however, these invariants do not provide a complete classification of, e.g., interacting topological phases \cite{FK10,HWML16,ZMBP23}.
Nevertheless, they have been widely used \cite{MENG12,HWGF13,HWML16,KMH16}, not least because, under suitable regularity conditions, they can be evaluated through the topological Hamiltonian \cite{WZ12,WY13,Gur11},
$\ff H_{\rm top}(\ff k) = - \ff G^{-1}(\ff k , i\omega = 0)$.
The topological Hamiltonian makes it possible to pose the same geometric question for interacting systems: 
beyond its conventional homotopy invariants, what additional topological structure is encoded in the geometry of $\ff H_{\rm top}(\ff k)$?

Green-function zeros provide a striking example of topological structure beyond conventional quasiparticle band descriptions.
In interacting systems, they arise naturally from singularities of the self-energy and can drive topological transitions alongside Green-function poles \cite{Gur11,WZ12}.
Recently, analogous zero-driven transitions have also been found in noninteracting systems with strong disorder \cite{VP26}.
After disorder averaging, translational invariance is restored and the resulting Green's function $\overline{\ff G}(\ff k,\omega)$ acquires a nontrivial self-energy, making disordered systems a very transparent setting for studying the geometry of Green-function-defined Hamiltonians without the additional complexity of many-body correlations.

\pp{Main result}
Here we pursue this idea for the disordered one-dimensional Su-Schrieffer-Heeger (SSH) model, with two sublattices, intracell and intercell hoppings $t_1$ and $t_2$, and disorder strength $W$.
Chiral symmetry of $\ff H_{\rm top}(k)=\ff d(k)\cdot\ff\sigma$, where $\ff\sigma$ denotes the Pauli matrices, constrains the Bloch vector $\ff d(k)$ to a plane.
As $k$ traverses the BZ, $\ff d(k)$ therefore traces a closed loop.
Besides its winding number $\nu$ around the origin, this loop possesses a second integer-valued quantity, its turning number $\tau$, which measures the winding of its {\em tangent vector} $\partial_k\ff d(k)$. 
The turning number is invariant under {\em regular} homotopy, i.e., under smooth deformations that preserve the immersion condition $\partial_k\ff d(k)\neq0$, unlike ordinary homotopy, where the relevant condition is $\ff d(k)\neq0$.

\begin{figure}[t] 
\centering
\includegraphics[width = 0.99\linewidth]{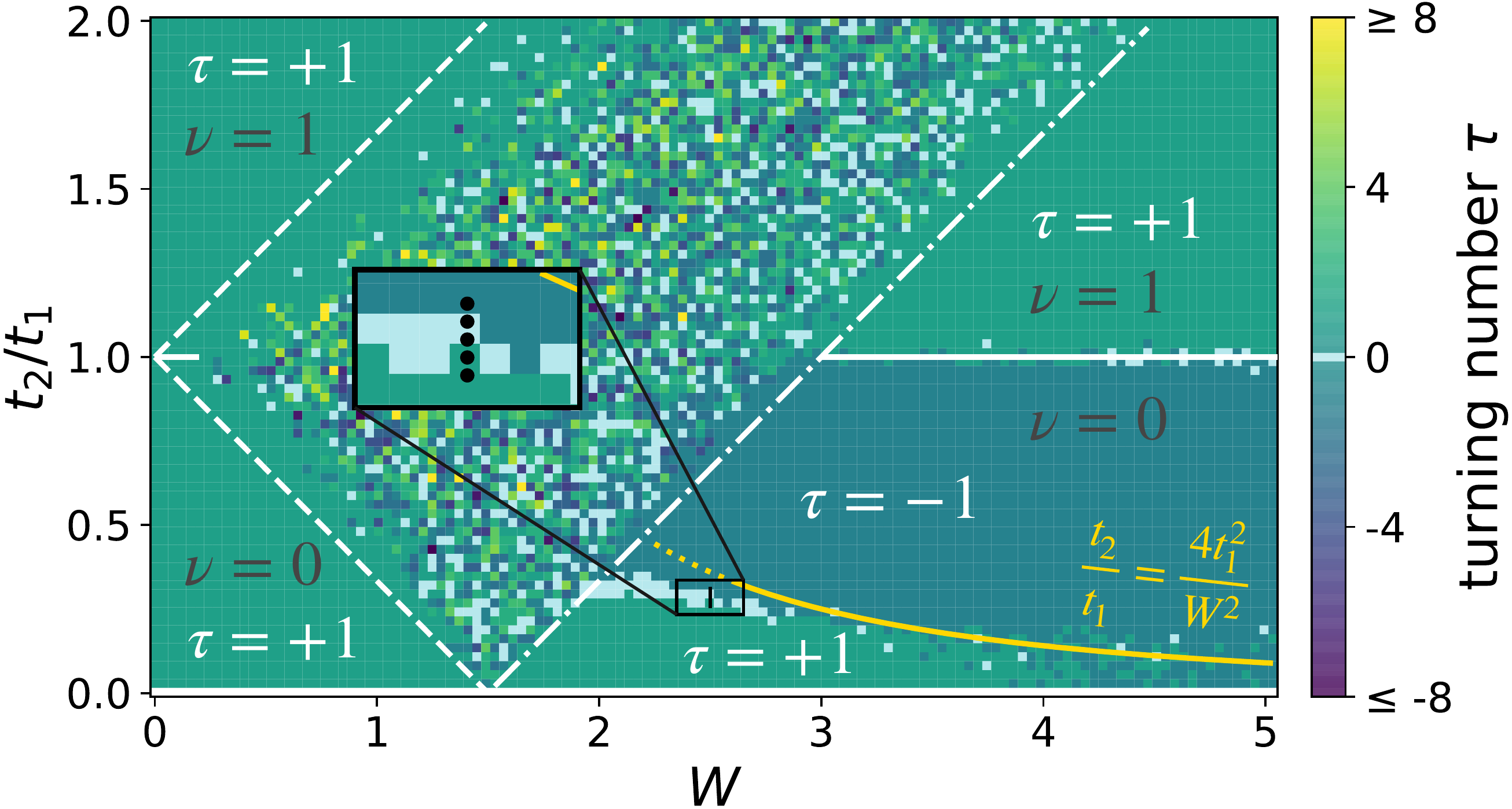}
\caption{
Turning number $\tau$ (color code) as a function of $t_{2}/t_{1}$ and $W$ for $t_{1}=0.75$.
Left of the dashed and right of the dashed-dotted line, Weyl's inequality guarantees an open spectral gap.
}
\label{fig:pd}
\end{figure}

Figure \ref{fig:pd} summarizes our central result.
While $\nu$ distinguishes the conventional topological sectors, $\tau$ reveals additional structure within them.
In the clean limit, the usual transition at $t_1=t_2$ is driven by a Green-function pole and changes $\nu$ from $0$ to $1$ without changing $\tau$.
At strong disorder, the corresponding transition along the same (solid white) line is instead driven by Green-function zeros; although $\nu$ again changes from $0$ to $1$, the two transitions are distinguished by $\tau$.
Moreover, $\tau$ exhibits an {\em additional} transition regime, located around the yellow line, within a fixed winding-number sector, revealing structure that is invisible to the conventional homotopy invariant.
Details of the computations, including an additional phase with $\tau=0$ (light blue) and the region with scattered $\tau$ values, are discussed below.
For the white dashed(-dotted) Weyl lines, see also Sec.\ A of the Supplemental Material (SM) \cite{sm}.

\begin{figure}[b] 
\centering
\includegraphics[width = 0.9\linewidth]{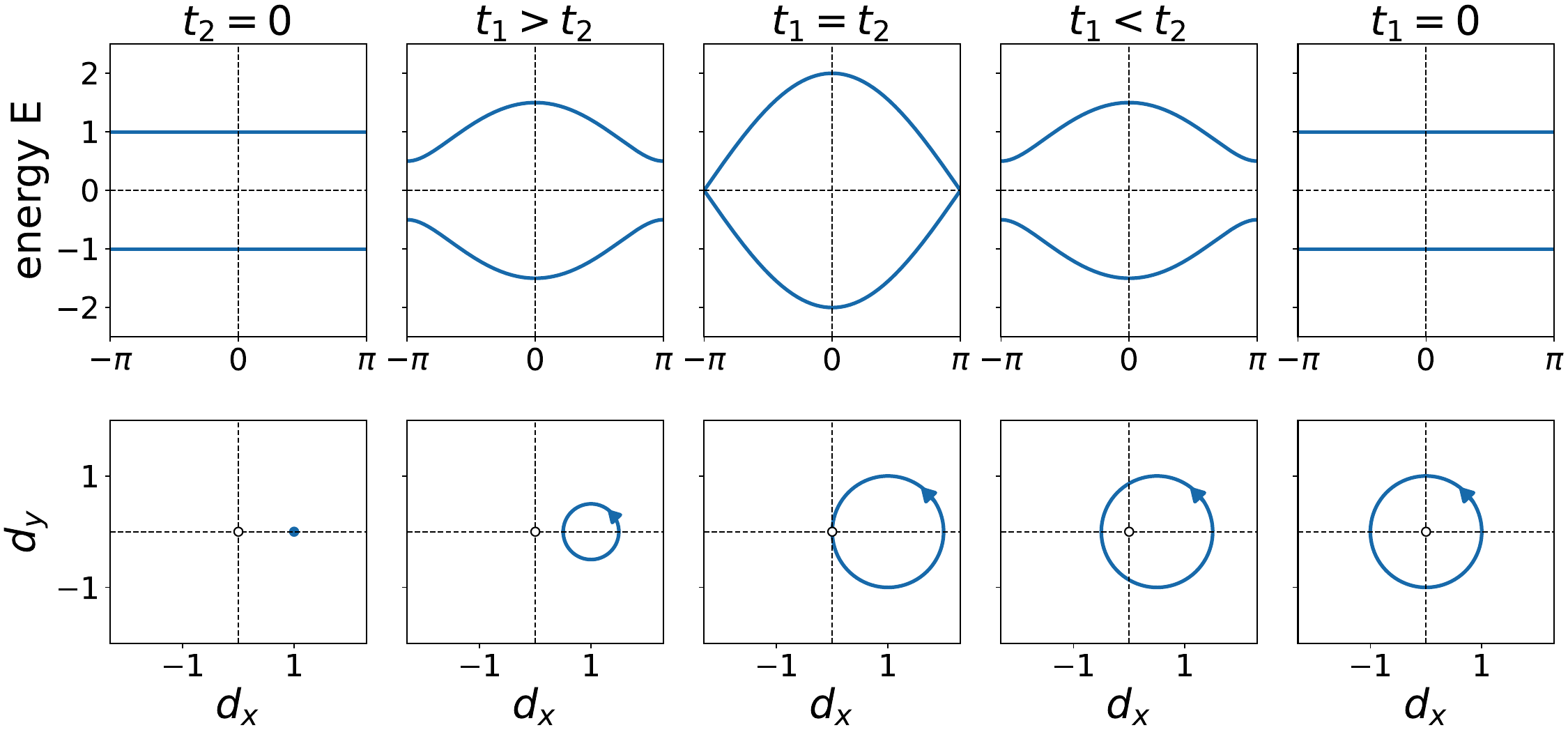}
\caption{
Band structure $\epsilon_{\pm}(k)$ for $-\pi < k \le \pi$ (top) and Bloch vector $\ff d_{0}(k)$ (bottom) in the clean limit $W=0$.
From left to right: $(t_{1},t_{2}) = (1,0), (1,0.5), (1,1), (0.5,1), (0,1)$. 
}
\label{fig:clean}
\end{figure}

\pp{Clean system} 
The clean SSH model \cite{SSH79} describes spinless fermions on a chain of $L$ unit cells with A and B sublattice sites:
$\hat{H}_{0} = \sum_{i=1}^{L} ( t_{1} a_{i}^{\dagger} b_{i} + t_{2} a^{\dagger}_{i+1} b_{i} + \mbox{H.c.} )$,
where $t_1$ and $t_2$ are the intra- and intercell hoppings.
Periodic boundary conditions ($a_{L+1} \equiv a_1$) are assumed.
For $t_2>t_1$, the system is in a chiral-symmetry-protected topological phase with winding number $\nu=1$, supporting localized zero-energy edge states under open boundary conditions.

The Bloch Hamiltonian reads
$\ff H_0(k)=\ff d_0(k)\cdot\ff\sigma$,
with $\ff d_0(k)=(t_1+t_2\cos k,t_2\sin k)$,
$\ff\sigma=(\sigma_x,\sigma_y)$, and $k\in(-\pi,\pi]$.
The Bloch vector defines a smooth map
$\ff d_0:S^1\to\mathbb R^2\setminus\{0\}$,
whose oriented winding around the origin is counted by $\nu\in\mathbb Z$.
Mathematically, $\nu$ labels the homotopy classes
$\pi_1(\mathbb R^2\setminus\{0\})\cong\mathbb Z$.
The loops and corresponding band structures are shown in Fig.\ \ref{fig:clean}.

The winding number becomes ill-defined when the loop crosses the origin, corresponding to a closing of the band separation
$\Delta(k)=2|\ff d_0(k)|=2\sqrt{t_1^2+t_2^2+2t_1t_2\cos k}$.
This occurs at $t_2=t_1$ and $k_{\rm c}=\pi$.
The Green's function
$\ff G_0(k,\omega)=(\omega-\ff H_0(k))^{-1}$
then satisfies
$\ff G_0(k_{\rm c},\omega)=\ff 1/\omega$,
implying a pole at $\omega=0$ and a divergent
$\det\ff G_0(k_{\rm c},\omega)$ as $\omega\to0$.

\pp{Adding disorder}
The winding number also admits a formulation directly in terms of the single-particle Green's function \cite{Gur11,MENG12}.
For a nonsingular Green's function at zero frequency, the winding number defined in this way equals that obtained from the topological Hamiltonian \cite{WZ12}.
It can change only when $\ff G(k,0)$ becomes singular, i.e., when $\det \ff G(k,0)$ vanishes or diverges for some $k$.
While poles already occur in clean noninteracting systems, Green-function zeros require interactions or, analogously, disorder averaging.

We consider binary potential disorder, i.e., $\hat{H} = \hat{H}_{0} + \hat{V}$ with
$\hat{V} = \sum_{i} (\epsilon_{i\rm A} a_{i}^{\dagger} a_{i} + \epsilon_{i\rm B} b_{i}^{\dagger} b_{i})$.
The variables $\epsilon_{i\rm A}$ and $\epsilon_{i\rm B}$ are independent and identically distributed, taking values $\pm W/2$ with equal probability.
$\hat{H}_{0}$ has chiral symmetry with respect to $\hat{\Gamma} = \mbox{diag}(\hat{1}_{\rm A} , -\hat{1}_{\rm B})$, i.e., the anticommutator $[ \hat{\Gamma} , \hat{H}_{0}]_{+} = 0$.
Although a generic disorder realization breaks chiral symmetry, we have
$\hat \Gamma ( \hat H_0 + \hat V ) \hat \Gamma = - ( \hat H_0 - \hat V)$.
Since $\hat V$ and $-\hat V$ occur with equal probability, the disorder-averaged Green's function recovers the chiral constraint at zero frequency.
Translational symmetry is likewise restored by disorder averaging.
For ease of notation, we denote its Fourier transform again by $\ff G(k,\omega)$.

\begin{figure}[t] 
\centering
\includegraphics[width = 0.9\linewidth]{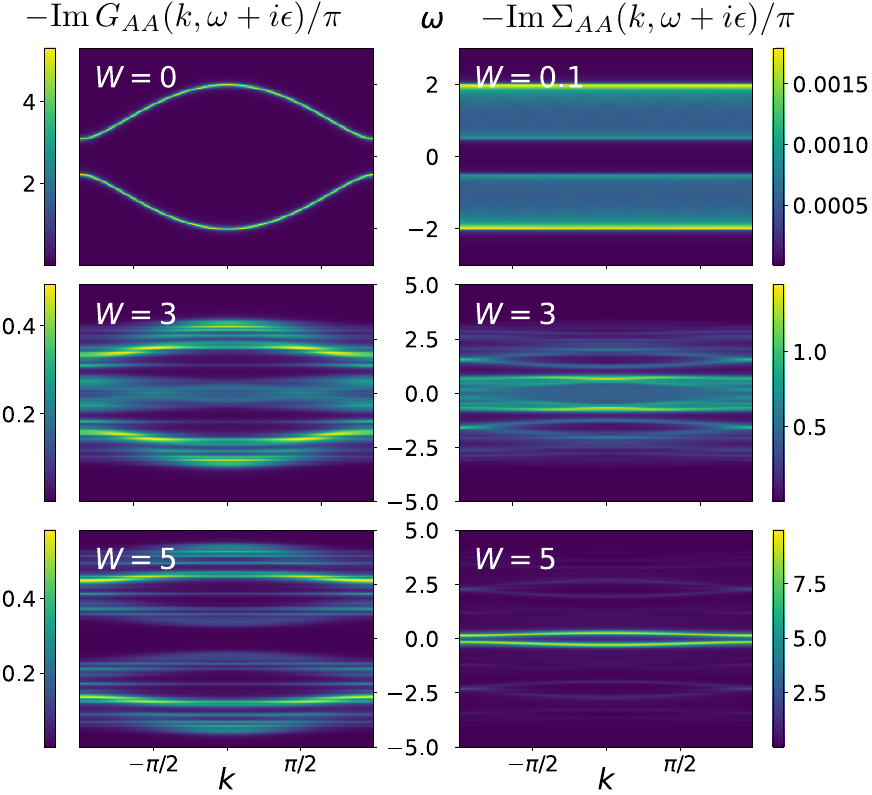}
\caption{
Imaginary part of the disorder-averaged Green's function (left) and of the disorder self-energy (right) 
for different disorder strengths $W$.
Parameters: $t_{1} = 0.75$, $t_{2} = 1.25$, $L = 400$, $N = 1000$. 
}
\label{fig:akw}
\end{figure}

\pp{Excitation spectum} 
Numerical results have been obtained by averaging over $N=1000$ independent configurations for a system with $L=400$ unit cells.
Figure \ref{fig:akw} (left) shows the evolution of the spectral density $A(k,\omega) \equiv A_{\rm AA}(k,\omega)=A_{\rm BB}(k,\omega)=(-1/\pi)\Im G_{\rm AA}(k,\omega+i0^{+})$ with increasing $W$, keeping $t_{1}=1.25$ and $t_{2}=0.75$ fixed.
The imaginary part of the retarded diagonal disorder self-energy $\Sigma_{\rm AA}(k,\omega+i0^{+})$ is shown in the right panels (note the different scales).
The high-frequency expansion yields the sum rule
$L^{-1} \sum_{k} \int_{-\infty}^{\infty}(-1/\pi) \mbox{Im} \Sigma_{\rm AA}(k,\omega + i 0^{+}) d \omega = W^{2} / 4$.

As seen in the figure, $A(k,\omega)$ undergoes a pronounced reconstruction.
The original clean bands at $W=0$ evolve into a number of weakly dispersive structures.
At $W=3$, these structures, together with their spectral weights, still loosely resemble the clean-limit dispersions, although they have been shifted towards higher absolute frequencies and strongly broadened.
At low frequencies around $\omega=0$, a broad low-energy spectral structure develops.
Upon further increasing $W$, this feature is progressively depleted around $\omega=0$, with the gap opening first near the BZ boundary and subsequently extending over the entire BZ.
At $W=5$, a full spectral gap has formed again.
The insulating state reached at large $W$, however, is qualitatively different from the clean $W=0$ insulator.
In particular, the reopening of the spectral gap is accompanied by two weakly dispersive branches of Green-function {\em zeros}, i.e., poles of the self-energy (lower-right panel of Fig.\ \ref{fig:akw}), at low frequencies.

\pp{Topology carried by zeros}
These zero bands in fact carry a nontrivial topology.
We have numerically computed the winding number via
$\nu = \frac{1}{2\pi} \int_{\rm BZ} dk \, \partial_k \arg  \left[ d_x(k) + i d_y(k) \right]$, 
where $\ff d(k)$ is obtained from $\ff H_{\rm top}(k) = \ff d(k) \cdot \ff \sigma = - \ff G^{-1}(k,\omega=0)$.
In the clean limit at $W=0$, $\nu$ jumps from $0$ to $1$ at $t_{2,\rm c} = t_{1} = 0.75$, where the gap closes at $k_{\rm c}=\pi$, see Fig.\ \ref{fig:clean}.
For strong disorder, e.g., $W=5$, the situation appears the same at the level of $\nu$:
The winding number again jumps from $\nu=0$ to $\nu=1$ at $t_{2,\rm c} = t_{1}$.
However, this transition is qualitatively different from the clean-limit case, since there is no gap closure at $t_{2,\rm c}$.
Rather, the two zero branches meet at $\omega=0$ and $k_{\rm c}=\pi$, i.e., $\det \ff G(k_{\rm c},\omega=0) = 0$ for $t_{2}=t_{2,\rm c}$.
Comparing $A(k,\omega)$ at $W=0$ with $(-1/\pi)\Im \Sigma_{\rm AA}(k,\omega)$ at $W=5$ for the noncritical value $t_{2}=1.25$, highlights the close analogy between the pole and zero structures, see the upper-left and lower-right panels of Fig.\ \ref{fig:akw}.
In Sec.\ B of the SM \cite{sm}, we demonstrate by a real-space calculation that the zero bands indeed carry the topology of the system.

This complements the findings of Wagner et al.\ \cite{WCA+23}, who demonstrated a Mott insulator whose nontrivial topology is carried by Green-function zeros, stimulating several related studies \cite{BF23,WGMS24,BSSK24,PPB25,FH25}; see also Refs.\ \cite{MN16,HNP22} and, for a two-dimensional disordered system, Ref.\ \cite{VP26}.
The numerical results for $\nu$ in the entire $W$-$t_{2}$ plane at fixed $t_{1}=0.75$ are shown in Sec.\ C of the SM.
The constant-$\nu$ phases are also indicated by the labels in Fig.\ \ref{fig:pd}.

\pp{Turning number} 
The winding number does not distinguish between the transitions at weak and strong $W$, i.e., it does not indicate whether poles or zeros carry the topology.
We therefore propose to additionally consider the turning number $\tau$ \cite{Whi37}, defined as the winding number of $\dot{\ff d}(k)\equiv \partial_k\ff d(k)$.
Thus, $\tau$ counts the windings of the {\em tangent vector} $\dot{\ff d}(k)$ rather than of $\ff d(k)$ itself.
Mathematically, $\tau=\deg ( \dot{\ff d} / |\dot{\ff d}| ) \in \mathbb{Z}$, provided that $\dot{\ff d}(k)\neq0$ throughout the BZ.
It is therefore a regular-homotopy invariant of the planar loop $\ff d(k)$ itself and, in particular, is independent of the position of the loop relative to the origin or any other fixed point.
Numerically, we evaluate
$\tau = \frac{1}{2\pi} \int_{\rm BZ} dk \, \partial_k \arg \left[ \partial_{k} d_x(k) + i \partial_{k} d_y(k) \right]$.
Compared with $\nu$, this amounts to replacing $\ff d(k)$ by $\partial_k\ff d(k)$.

Applications of the turning number have so far mainly concerned the geometry of immersed momentum-space contours, such as figure-eight Fermi surfaces in thin-film Weyl semimetals \cite{BBB+18}, spiral Fermi surfaces in quasicrystalline systems and twisted bilayer graphene \cite{SC19}, and higher-turning-number spiral Fermi surfaces in twisted multilayer systems \cite{WH25}. 
Turning-number-based immersion classes have also been used to construct superselection sectors for anyonic systems \cite{Shi26}.
By contrast, here the turning number characterizes the geometry of the map $k\mapsto\ff H_{\rm top}(k)$.

In the clean limit $W=0$, see Fig.\ \ref{fig:clean}, $\tau$ is insensitive to the topological transition.
Upon varying $t_2/t_1$, the planar loop $\ff d_{0}(k)$ passes through the origin $\ff d_{0}=0$ at $t_{2,\rm c}=t_1$, where $\nu$ changes from $\nu=0$ to $\nu=1$.
The tangent vector, however, remains nonzero throughout, including at $t_{2,\rm c}$.
As the BZ is traversed once in the positive direction, the tangent vector winds once in the positive direction.
Consequently, $\tau=+1$ independently of $t_2/t_1$.

\begin{figure}[t] 
\centering
\includegraphics[width = 0.99\linewidth]{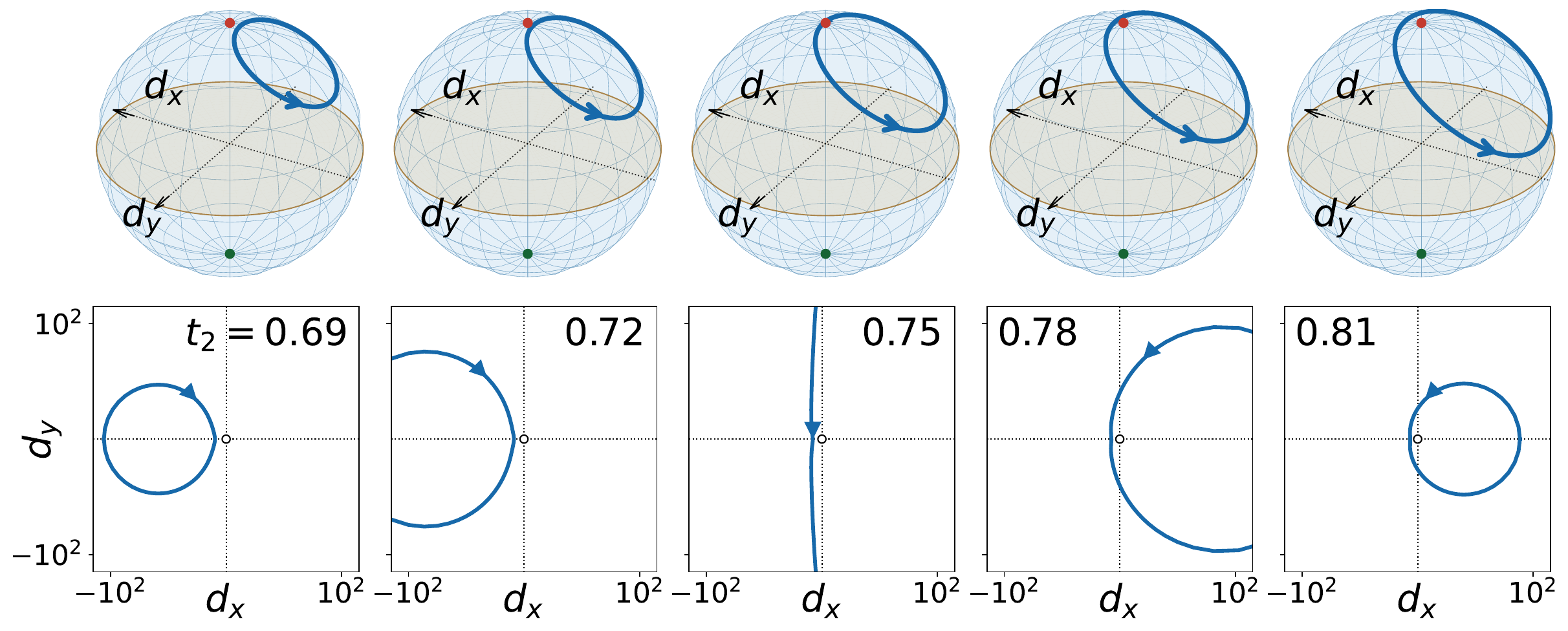}
\includegraphics[width = 0.99\linewidth]{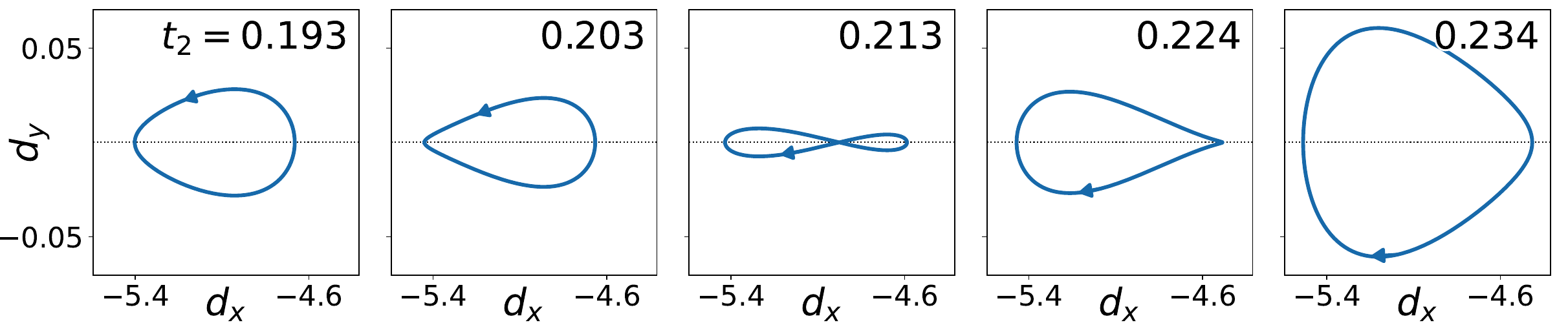}
\caption{
Loop $\ff d(k)$ in the $d_{x}$-$d_{y}$ plane (middle) and its inverse stereographic projection onto the sphere $S^{2}$ (top panels) for various $t_{2}$ at $W=3.5$
Bottom: The same but for $W=2.5$. 
Parameters: $t_{1}=0.75$, $L=600$, $N=1000$. 
}
\label{fig:zero}
\end{figure}

For strong disorder $W=3.5$, by contrast, the loop passes through ``infinity'' rather than through $\ff d=0$.
This is visualized in Fig.\ \ref{fig:zero} (top panels) using an inverse stereographic projection onto a sphere representing the compactified $\ff d$ plane.
The north pole (N, red dot) and south pole (S, green dot) represent $|\ff d|=\infty$ and $\ff d=0$, respectively.
Exactly at $t_{2}=t_{1}$, the loop reaches N at $k=\pi$.
The middle panels in Fig.\ \ref{fig:zero} display the evolution of the loop in the $\ff d$ plane.
With increasing $t_{2}$, but still $t_{2}<t_{1}$, the loop grows.
For $t_{2}\to t_{2,\rm c}=t_{1}$, the loop becomes singular and extends to infinity, with $d_x(\pi)\to-\infty$, since a zero of $\det\ff G(k,\omega)$ crosses $\omega=0$ at $k=\pi$.
For $t_{2}>t_{1}$, the loop reappears from infinity on the opposite ($d_x>0$) side and closes again; upon further increasing $t_{2}$, it shrinks.
Importantly, the loops for $t_{2}<t_{1}$ and $t_{2}>t_{1}$ differ in both invariants: 
the winding number jumps from $\nu=0$ to $\nu=1$, while the turning number changes from $\tau=-1$ to $\tau=+1$.
The latter qualitatively distinguishes the strong-disorder transition from its clean-limit counterpart and, for $t_{2}<t_{1}$, also distinguishes the clean and strong-disorder regimes despite their common winding number $\nu=0$.

\pp{Refined topological phase diagram}
Figure \ref{fig:pd} shows the entire $W$-$t_{2}$ topological phase diagram at fixed $t_{1}=0.75$ as obtained for $L=200$ unit cells and $N=2000$ disorder configurations (see also SM \cite{sm}, Sec.\ C).
The winding and turning numbers have been obtained numerically as described above.
There is a parameter region where $\nu$ and $\tau$ are ill-defined as topological invariants, since the disorder-averaged spectral function has finite weight at zero frequency, $A(k,0)>0$ for some $k$, and the zero-frequency retarded Green's function, and hence $\ff H_{\rm top}(k)$, is no longer Hermitian.
The numerical winding formulas nevertheless still return integer values.
These values, however, are no longer topologically protected and fluctuate strongly between neighboring points in parameter space, rather than forming extended regions of constant $\nu$ and $\tau$.
Weyl's inequality \cite{HJ13} provides a sufficient criterion for an open spectral gap and hence for $\nu$ to be well defined; see SM \cite{sm}, Sec.\ A, and the white dashed and dashed-dotted lines in Fig.\ \ref{fig:pd}.

Outside the region bounded by the Weyl lines, we find $\nu=0$ for $t_{2} / t_{1}<1$ and $\nu=1$ for $t_{2} / t_{1}>1$.
The transition is pinned to $t_2=t_1$ by the symmetry of the disorder-averaged problem under interchange of the intra- and intercell hoppings, $t_1\leftrightarrow t_2$ (SM \cite{sm}, Sec.\ D).
The turning number, by contrast, is not constrained by this hopping-exchange symmetry, and its transitions can occur away from $t_2=t_1$.

While the winding number $\nu$ is the topological invariant characterizing one-dimensional chiral systems in class AIII, the turning number $\tau$ classifies the corresponding planar loops of the topological Hamiltonian under regular homotopy, i.e., under continuous deformations for which the tangent vector never vanishes (the ``immersion'' condition).
It therefore resolves additional geometric structure within a fixed winding-number sector.

Changes of $\tau$ between gapped regions can occur across a gapless region, as for $t_{2}<t_{1}$, or across a divergence of $\ff H_{\rm top}$ associated with a Green-function zero at $t_{2}=t_{1}$.
More interestingly, however, $\tau$ can also change while $\ff H_{\rm top}$ remains finite and gapped, namely when the immersion condition is lost, $\partial_k\ff d(k)=0$ for some $k$.
Such a transition has no counterpart in the winding-number classification.

\pp{Immersion-driven turning-number transition}
In fact, such a transition is predicted by a perturbative calculation of the complexified $d$ vector, $d_{+} \equiv d_{x}+i d_{y}$, for weak $t_{2} \to 0$ and strong $W \to \infty$.
The corresponding asymptotic expansions depend on the order in which these limits are taken.
Expanding first in $t_{2}/t_{1}$ and subsequently taking $W \gg t_{1}$, one finds
$d_+(k) \simeq -\frac{W^4}{16t_1^3} + t_2 e^{ik}$.
On the other hand, taking $W \to \infty$ first and subsequently expanding in $t_2/t_1$ yields, for $t_2<t_1$,
$d_+(k) \simeq -\frac{W^4}{16t_1^3} + \frac{W^4t_2^3}{16t_1^6} e^{-ik}$ (see SM \cite{sm}, Secs.\ E and F for details).
We note that the respective $k$-independent leading terms are identical.
Importantly, however, the leading $k$-dependent corrections have opposite windings.
The crossover between the two asymptotic regimes is therefore expected when the amplitudes of these corrections become comparable.
Equating the corresponding coefficients gives the crossover scale
$t_{2,\rm c}\simeq\frac{4t_1^3}{W^2}$.
This provides an estimate for the onset of the turning-number transition regime in the $W$-$t_2$ phase diagram, indicated by the yellow line in Fig.\ \ref{fig:pd}.

The agreement with the numerically obtained $\tau$ values is nearly quantitative for $W \gtrsim 3$, see Fig.\ \ref{fig:pd}.
For $W \gtrsim4 $, however, the numerical accuracy deteriorates, leading to scattered data.
Since perturbation theory, at least to the orders considered, cannot distinguish between a direct $\tau=-1\to+1$ transition and a transition via an intermediate $\tau=0$ phase, we performed $t_{2}$-dependent calculations at fixed $W=2.5$ (see black dots in the inset in Fig.\ \ref{fig:pd}) for larger systems with $L=600$ unit cells.
The bottom panels of Fig.\ \ref{fig:zero} indeed demonstrate the existence of a $\tau=0$ (figure-eight) phase for $t_{2,\rm c1} \approx 0.203 < t_{2} < 0.224 \approx t_{2,\rm c2}$.
At the two critical values, the $\ff d$ loops exhibit cusp singularities at the time-reversal-invariant momenta $k=0$ and $\pi$, respectively (see also SM \cite{sm}, Sec.\ G).
Altogether, these results show that the regular-homotopy classification by $\tau$ reveals geometric distinctions of the topological Hamiltonian that are invisible to its conventional homotopy classification by $\nu$.

\pp{Perspectives}
Our results suggest several immediate directions.
A particularly important one is to search for analogous $\tau$-resolved phases and immersion-driven transitions in interacting one-dimensional systems, thereby testing whether the geometric refinement found here extends beyond disorder.
A complementary direction is to develop the gauge-covariant formulation of this geometric refinement, in which the immersion condition can be expressed in terms of a covariant derivative of $\ff H_{\rm top}$.
Another direction concerns a bulk-boundary correspondence for $\tau$, for example through the spectrum of $\partial_k\ff H_{\rm top}$ or a related current operator.
The immersion transition is expected to leave observable signatures, as the associated change in the geometry of the Hamiltonian map can modify van Hove singularities and thereby produce characteristic features in optical response.
Most importantly, the present construction motivates geometric classifications in higher dimensions, with immersion singularities providing a natural starting point for two-dimensional Hamiltonian maps.

\pp{Acknowledgments} 
This work was supported by the Deutsche Forschungsgemeinschaft (DFG, German Research Foundation) through the research unit QUAST, FOR 5249 (project P8), project ID 449872909.

\begin{widetext}
\newpage
\mbox{}
\setcounter{page}{1}

\begin{center}
{\large \bfseries 
\titlepaper

\mbox{}\\

--- Supplemental Material ---
}

\mbox{}\\

{
Lars Emmrich and Michael Potthoff
}

\mbox{}\\[-2mm]

{
\small \it 
Department of Physics, University of Hamburg, Notkestra{\ss}e 9-11, 22607 Hamburg, Germany
}

\end{center}

\end{widetext}

\mbox{} \newline \pp{Section A: Implications from Weyl's inequality}  \newline

\noindent
Weyl's inequality \cite{HJ13} provides a simple rigorous bound how disorder can reduce the gap of the SSH model.
Let $\hat{A} = \hat{A}_{0} + \hat{A}_{1}$, where $\hat{A}_{0}$ and $\hat{A}_{1}$ are Hermitian operators acting on a linear space of finite dimension $D$.
Weyl's inequality implies
\be
\left| \lambda_n (\hat{A}) - \lambda_n (\hat{A}_0) \right|
\le
\| \hat{A}_{1} \|
\: , 
\labeq{we}
\ee
where $\lambda_n(\hat{A})$ denote the eigenvalues of a Hermitian operator $A$, ordered as 
$\lambda_{1}(A) \ge \cdots \ge \lambda_{n}(A) \ge \cdots \ge \lambda_{D}(A)$. 
Furthermore, $\| \hat{A} \| = \sup_{\|\psi\|=1}\| \hat{A} \psi \|$ is the operator norm of $\hat{A}$.
Equivalently, $\| \hat{A} \| = \max_n | \lambda_{n}(A) |$. 

Let $\hat{H} = \hat{H}_{0} + \hat{V}$, where $\hat{H}_{0}$ and $\hat{V}$ are the clean Hamiltonian and the disorder potential, respectively. 
Here, for binary disorder, we have $\| \hat V \| = W/2$. 
With \refeq{we} we can conclude that the unoccupied states have energies $E_{n}$ with
\be
E_{n} \ge \frac{\Delta}{2} - \| \hat V \| = \frac{\Delta}{2} - \frac{W}{2}
\: ,
\ee
where $\Delta$ denotes the gap in the clean limit.
Analogously, the occupied states have energies such that
\be
E_{n} \le - \frac{\Delta}{2} + \| \hat V \| = - \frac{\Delta}{2} + \frac{W}{2}
\: .
\ee
Hence the disordered system has a spectral gap satisfying
\be
\Delta(W) \ge \Delta - W
\: .
\ee
In the clean limit, $\hat{H}_{0}$, the gap is given by
\be
\Delta
=
2
\big|
|t_1|-|t_2|
\big|
\: .
\ee
Thus
\be
\Delta(W) \ge 2 \bigl|| t_1 | - | t_2 | \bigr| - W
\: .
\ee
In particular, the gap is guaranteed to remain open as long as
\be
W < 2 \bigl| |t_1| - |t_2| \bigr|
\: .
\labeq{w1}
\ee

One may apply \refeq{we} once more, but with exchanged roles of $\hat{H}_{0}$ and $\hat{V}$.
This is meaningful in the strong-disorder regime $W \gg |t_{1}| , |t_{2}|$.

The ``unperturbed'' operator $\hat{V}$ has only the two eigenvalues $\pm W/2$.
Treating $\hat{H}_0$ as the perturbation, Weyl's inequality, see \refeq{we}, implies that every eigenvalue of $\hat{H}_{0}$ can be shifted by at most $\| \hat{H}_{0} \|$.
Consequently, the lower group of eigenvalues of $\hat{H}$ is contained in the interval
\be
\left[
-\frac{W}{2} - \| \hat{H}_{0} \|, 
-\frac{W}{2} + \| \hat{H}_{0} \|
\right]
\: ,
\ee
whereas the upper group is contained in
\be
\left[
+ \frac{W}{2} - \| \hat{H}_{0} \|,
+ \frac{W}{2} + \| \hat{H}_{0} \|
\right]
\: .
\ee
Hence the gap between the two spectral regions satisfies
\be
\Delta(W)
\geq
W - 2 \| \hat{H}_{0} \|
\: .
\ee
A finite gap is therefore guaranteed whenever
\be
W > 2 \| \hat{H}_{0} \|
\: .
\ee

For the clean SSH Hamiltonian with intracell and intercell hoppings $t_1$ and $t_2$, the band dispersions are given by 
\be
\epsilon_{\pm}(k)
=
\pm \left| t_1 + t_2 e^{-ik} \right|
\: , 
\ee
and thus
\be
\| \hat{H}_{0} \| = \max_k \left| t_1 + t_2 e^{-ik} \right | = | t_1 | + | t_2 |
\: .
\ee
Therewith, we find 
\be
\Delta(W)
\ge
W - 2 \left( |t_1| + |t_2| \right)
\: , 
\ee
and thus
\be
W > 2 \left( |t_1| + |t_2| \right)
\labeq{w2}
\ee
is a sufficient condition for the spectrum to be gapped.

It should be stressed that the Weyl bounds give sufficient conditions for the persistence of the spectral gap, but not the actual critical disorder strengths. 
We can combine \refeq{w1} and \refeq{w2}: 
If
\be
W <
2\bigl||t_1|-|t_2|\bigr|
\labeq{g1}
\ee
the gap is certainly open. 
If
$
2\bigl||t_1|-|t_2|\bigr|
\leq W \leq
2\left(|t_1|+|t_2|\right)
$
Weyl's inequality gives no conclusion. 
If 
\be
W >
2\left(|t_1|+|t_2|\right)
\labeq{g2}
\ee
the gap is certainly open again.

\mbox{} \newline \pp{Section B: Edge poles and edge zeros}  \newline

\noindent
In the strong-disorder regime and for $t_{2}>t_{1}$, the imaginary part of the disorder self-energy,
$(-1/\pi)\Im\Sigma_{\rm AA}(k,\omega+i\epsilon)$, exhibits two weakly dispersing sharp features in the low-frequency regime, as shown in the bottom-right panel of Fig.\ \ref{fig:akw}.
The apparent finite spectral width of these features results exclusively from the finite broadening parameter.
They therefore correspond to poles of the self-energy and, consequently, to zero bands of the Green's function $G(k,\omega)$.
In the parameter regime considered here, the winding number obtained from the topological Hamiltonian is nontrivial and given by $\nu=1$.

\begin{figure}[t]
\centering
\includegraphics[width = 0.7\linewidth]{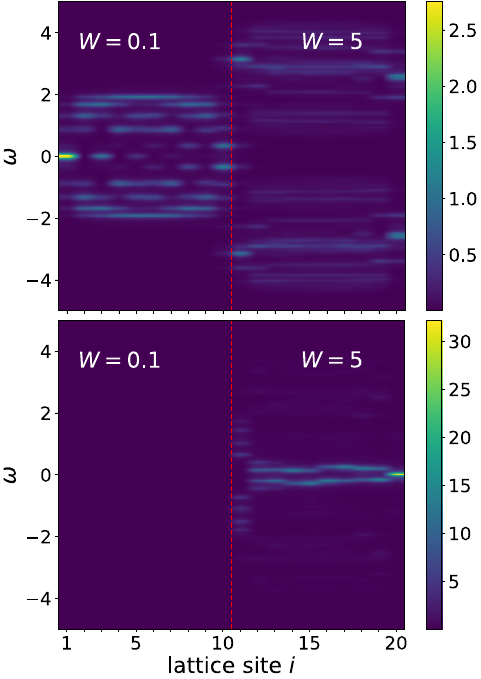}
\caption{
Top: Imaginary part of the local, disorder-averaged Green's function,
$(-1/\pi)\Im G_{ii}(\omega+i\epsilon)$, with $i=1,\ldots,20$ labeling the sites of an SSH chain with disorder strength $W=0.1$ in the left half and $W=5$ in the right half (left and right of the dashed red line).
Bottom: Imaginary part of the corresponding disorder self-energy,
$(-1/\pi)\Im\Sigma_{ii}(\omega+i\epsilon)$.
Parameters: $t_{1}=0.75$, $t_{2}=1.25$, $L=20$, $N=1000$, $N_{\omega}=200$, and $\epsilon=\Delta\omega$.
}
\label{fig:edge}
\end{figure}

The fact that this nontrivial topology is carried by the zero bands can be further corroborated by considering the disordered SSH model in real space with open boundaries, as shown in Fig.\ \ref{fig:edge}.
Specifically, we consider a chain of $L=20$ sites consisting of two halves: an almost clean part on the left, with $W=0.1$ on sites $i=1,\ldots,10$, and a strongly disordered part on the right, with $W=5$ on sites $i=11,\ldots,20$.
The sites $i=1$ and $i=20$ form the outer boundaries and are coupled to their respective neighboring sites by the weaker hopping $t_{1}<t_{2}$.

On the left-hand side, the nontrivial topology is represented by poles of the Green's function.
For $\nu=1$, the bulk-boundary correspondence predicts a zero-energy edge pole.
This is clearly visible in the top panel of Fig.\ \ref{fig:edge} at the left boundary, $i=1$, with a spectral weight that rapidly decreases with increasing distance from the edge.
By contrast, no corresponding low-energy spectral weight is found at the interface site $i=10$.

Conversely, on the strongly disordered right-hand side, the nontrivial topology is represented by an edge zero.
It appears as a pole of the self-energy with large spectral weight at the right boundary, $i=20$, in the lower panel of Fig.\ \ref{fig:edge}.
Again, the weight rapidly decreases with increasing distance from the boundary, while no corresponding low-energy self-energy peak is found at the interface site $i=11$.

To make the interface physics explicit, we denote the hopping between sites $i=10$ and $i=11$ by $t_{\rm int}$.
Upon switching off this hopping, $t_{\rm int}=0$, a second edge pole appears on the left part at $i=10$ (not shown), while a second edge zero appears on the right part at $i=11$ (not shown), as expected from the bulk-boundary correspondence for the two topologically nontrivial subsystems.
Upon restoring $t_{\rm int}$ to its bulk value, the pole and zero at the interface hybridize and the zero-energy interface singularities disappear.
This realizes the ``topological matter-antimatter annihilation'' \cite{WCA+23}, i.e., a Green-function pole annihilates with a Green-function zero.
The effect is analogous to the pole-zero annihilation found in Ref.\ \cite{WCA+23} at the interface between a noninteracting topological SSH chain and a strongly interacting topological SSH+$U$ chain.

\mbox{} \newline \pp{Section C: Topological phase diagram obtained from the winding number}  \newline

\begin{figure}[t] 
\centering
\includegraphics[width = 0.95\linewidth]{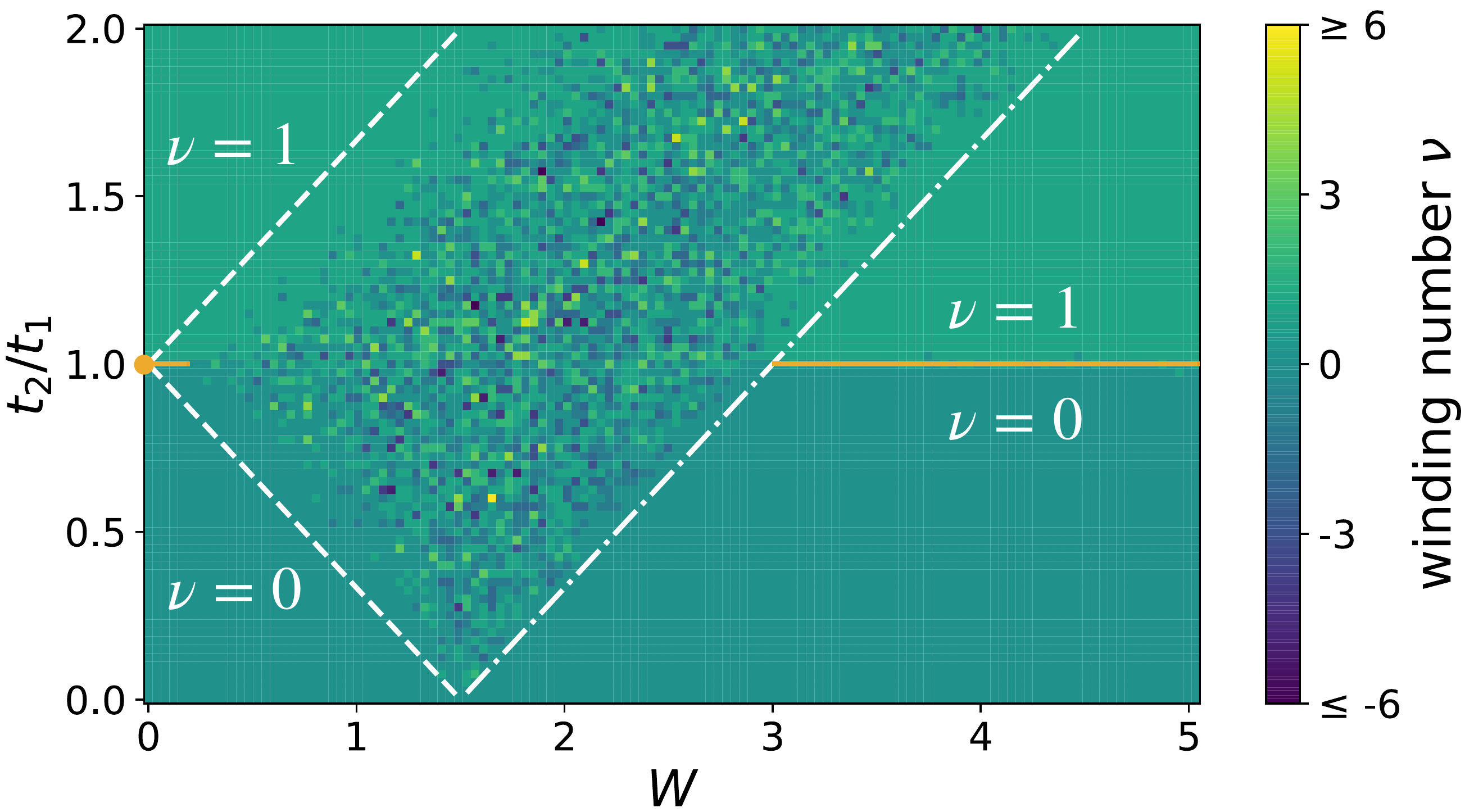}
\caption{
Winding number $\nu$ (color code) as function of $t_{2}/t_{1}$ and $W$ for $t_{1} = 0.75$. 
Orange dot: critical point at $W=0$.
Orange line: topological transition line.
Left of the dashed and right of the dashed-dotted white lines, Weyl's inequality guarantees that the spectral gap is open, see Eqs.\ (\ref{eq:g1}) and (\ref{eq:g2}).
Parameters: $L=200$, $N=2000$.
}
\label{fig:pdwinding}
\end{figure}

Figure \ref{fig:pdwinding} shows the result of a numerical computation of the winding number $\nu$, as defined via $\nu = \frac{1}{2\pi} \int_{\rm BZ} dk \, \partial_k \arg  \left[ d_x(k) + i d_y(k) \right]$, in the $W$-$t_{2}$ plane for fixed $t_{1}=0.75$. 
The phases with $\nu = 0$ for $t_{2} < t_{1}$ and $\nu=1$ for $t_{2}>t_{1}$ in the clean limit (see orange dot) or at weak disorder are clearly separated from $\nu = 0$ and $\nu=1$ phases in the strong-disorder limit by a gapless phase. 

The exact boundaries of the gapless phase are difficult to determine numerically. 
However, Weyl's inequality \cite{HJ13} is very useful to identify the region, where the gap is certainly open and thus where $\nu$ is well-defined as a topological invariant. 
This is discussed in Sec.\ A of the SM, see Eqs.\ (\ref{eq:g1}) and (\ref{eq:g2}) in particular.
Left of the dashed and right of the dashed-dotted line, Weyl's inequality guarantees that the spectral gap is open. 

Formally, the winding of an individual closed $\mathbf d(k)$ loop, as computed from 
$\nu = \frac{1}{2\pi} \int_{\rm BZ} dk \, \partial_k \arg  \left[ d_x(k) + i d_y(k) \right]$, remains integer-valued as long as the loop avoids the origin. 
However, for a gapless phase, these integers are not topologically protected and fluctuate strongly between neighboring points in parameter space, rather than forming extended regions of constant $\nu$.

\mbox{} \newline \pp{Section D: Chiral symmetry and the $t_{1}=t_{2}$ line}  \newline

\noindent
The clean SSH Hamiltonian possesses (sublattice) chiral symmetry,
\be
\ff \Gamma \ff H_{0}(k) \ff \Gamma = - \ff H_{0}(k)
\: , 
\labeq{gammah}
\ee
where the chiral matrix $\ff \Gamma$ is unitary and $\ff \Gamma^2=\ff 1$.
While individual disorder configurations generically break chiral symmetry, the symmetry is restored by the disorder ensemble, and hence
\be
\ff \Gamma \ff G(k,\omega) \ff \Gamma = - \ff G(k,-\omega)
\labeq{gamma}
\ee
for the disorder-averaged Green's function at complex frequencies.
We note in passing, that this immediately implies
\be
\ff \Gamma \ff H_{\rm top}(k) \ff \Gamma = - \ff H_{\rm top}(k)
\labeq{chir}
\ee
for the topological Hamiltonian $\ff H_{\rm top}(k) = - \ff G^{-1}(k , i\omega = 0)$.

Taking the determinant of the $2 \times 2$ matrices on both sides of \refeq{gamma}, and using $\ff \Gamma^2=\ff 1$, we get
\be
\det \ff G(k,\omega) = \det \ff G(k,-\omega)
\: .
\ee
Now let $\omega_{z,n}(k)$ denote a Green-function zero,
$\det \ff G\bigl( k,\omega_{z,n}(k) \bigr) = 0$, then
\be
0
=
\det \ff G\bigl(k,\omega_{z,n}(k)\bigr)
=
\det \ff G\bigl(k,-\omega_{z,n}(k)\bigr)
\: .
\ee
Hence, for every zero at positive frequency there must be a zero at the corresponding negative frequency at the same momentum:
\be
\omega_{z,\bar n}(k)=-\omega_{z,n}(k)
\: .
\ee
For two zero branches, labeled by $+$ and $-$, this reads
\be
\omega_{z,+}(k)=-\omega_{z,-}(k)
\: .
\ee
Analogously, we have 
\be
\omega_{p,+}(k)=-\omega_{p,-}(k)
\: 
\ee
for the Green-function poles.
Consequently, if one of the two chiral-symmetry-related zero bands reaches the Fermi level at any momentum $k_c$, i.e., $\omega_{z,+}(k_c)=0$, then its partner necessarily reaches the Fermi level at the same momentum:
\be
\omega_{z,+}(k_c)
=
\omega_{z,-}(k_c)
=
0
\: .
\ee

We now consider the unitary transformation $\hat U$ acting on the annihilators as 
\be
a_{i} \mapsto \hat U a_{i} \hat U^{\dagger} = b_{i}
\; , \qquad 
b_{i} \mapsto \hat U b_{i} \hat U^{\dagger} = a_{i+1}
\: . 
\ee
Unter $\hat U$, the clean Hamiltonian, with periodic boundary conditions $a_{L+1} \equiv a_{1}$, transforms as
\ba
\hat H_{0} &=& \sum_{i=1}^{L} ( t_{1} a_{i}^{\dagger} b_{i} + t_{2} a^{\dagger}_{i+1} b_{i} + \mbox{H.c.} )
\nonumber \\
&\mapsto&
\hat U \hat{H}_{0} \hat U
= 
\sum_{i=1}^{L} ( t_{1} b_{i}^{\dagger} a_{i+1} + t_{2} b^{\dagger}_{i+1} a_{i+1} + \mbox{H.c.} )
\nonumber \\
&=& 
\sum_{i=1}^{L} ( t_{2} b^{\dagger}_{i} a_{i} + t_{1} b_{i}^{\dagger} a_{i+1} + \mbox{H.c.} )
\nonumber \\
&=& 
\sum_{i=1}^{L} ( t_{2} a^{\dagger}_{i} b_{i} + t_{1} a_{i+1}^{\dagger} b_{i} + \mbox{H.c.} )
\ea
with $b_{L+1} \equiv b_{1}$.
Hence $\hat U \hat H_{0}(t_{1},t_{2}) \hat U^{\dagger} = \hat H_{0}(t_{2},t_{1})$, i.e., intra- and intercell hopping amplitudes are interchanged.

In momentum space, 
\be
\hat H_{0} = \sum_{k\in \rm BZ} 
\begin{pmatrix} a^\dagger_{k} , b^{\dagger}_{k} \end{pmatrix}
\ff H_{0}(k) 
\begin{pmatrix} a_{k} \\ b_{k} \end{pmatrix}
\: , 
\ee
with 
\be
\ff H_{0}(k)
=
\begin{pmatrix}
0 & t_{1} + t_{2} e^{-ik} \\
t_{1} + t_{2} e^{ik} & 0 \\
\end{pmatrix}
\; , 
\ee
the transformation acts as 
\be
\ff H_{0}(k) \mapsto \ff U(k) \ff H_{0}(k) \ff U(k)^\dagger 
\: , \; \mbox{with} \quad 
\ff U(k)
=
\begin{pmatrix}
0 & 1 \\
e^{ik} & 0 \\
\end{pmatrix}
\: , 
\ee
with the analogous result
\be
\ff U(k) \ff H_{0}(t_{1},t_{2};k) \ff U(k)^\dagger = \ff H_{0}(t_{2},t_{1};k)
\: .
\ee

The effect of the unitary transformation on the disorder potential is 
\ba
\hat{V} &=& \sum_{i} ( \epsilon_{i\rm A} a_{i}^{\dagger} a_{i} + \epsilon_{i\rm B} b_{i}^{\dagger} b_{i})
\nonumber \\
&\mapsto& \sum_{i} ( \epsilon_{i\rm A} b_{i}^{\dagger} b_{i} + \epsilon_{i\rm B} a_{i+1}^{\dagger} a_{i+1})
\: .
\ea
Since the parameters $\epsilon_{i\rm A}$, $\epsilon_{i\rm B}$ for all $i$ are independent and identically distributed, 
the disorder ensemble is invariant.

For the disorder-averaged Green's function we thus have 
\be 
\overline{\ff G}(t_2,t_1; k,\omega)
=
\ff U(k) \,
\overline{\ff G}(t_1,t_2; k,\omega) \,
\ff U^\dagger(k)
\: ,
\ee
and therefore
\be 
\ff H_{\rm top}(t_2,t_1; k)
=
\ff U(k) \,
\ff H_{\rm top}(t_1,t_2;k) \,
\ff U^\dagger(k)
\: . 
\labeq{tauhk}
\ee
Chiral symmetry, \refeq{chir}, implies $\ff H_{\rm top}(k) = d_{x}(k) \ff \sigma_{x} +d_{y}(k) \ff \sigma_{y}$: 
\be 
\ff H_{\rm top}(k)
=
\begin{pmatrix}
0 & q^{\ast}(k) \\
q(k) & 0 \\
\end{pmatrix}
\ee
with $q(k)=d_{x}(k) + id_{y}(k)$.
With \refeq{tauhk}, this yields
\be
q(t_{2},t_{1};k) = e^{ik} q^{\ast}(t_{1},t_{2};k) 
\: .
\labeq{qwind}
\ee
The winding number can be expressed in terms of $q(k)$ as 
\be
\nu(t_{1},t_{2}) = \frac{1}{2\pi i} \int_{\rm BZ} dk \, \partial_{k} \ln q(t_{1},t_{2};k) 
\: .
\ee
Since $k \mapsto e^{ik}$ has winding number $+1$ and since complex conjugation reverses the winding, \refeq{qwind} implies
\be
\nu(t_{2}, t_{1}) = 1 - \nu(t_{1},t_{2})
\: .
\labeq{nusym}
\ee
There is no analogous constraint for the turning number, since differentiating 
\refeq{tauhk} with respect to $k$ produces an extra term due to the $k$ dependence of the unitary transformation. 
Hence, there need not be a symmetry relation for the $\tau$ phase diagram with respect to exchanging $t_{1}$ with $t_{2}$. 

Note that both the winding and the turning number are evaluated in a fixed real-space orbital representation of the SSH model. The unitary transformation considered here relates different points of the parameter space in this same representation, rather than serving as a passive change of orbital basis.

We also note that $\nu$ and $\tau$ are evaluated in the natural real-space orbital representation of the SSH model, with a fixed unit-cell convention and real hopping amplitudes. 
Under a momentum-dependent change of orbital basis, $\widetilde{\ff H}(k) =\ff U(k) \ff H(k) \ff U^\dagger(k)$, ordinary $k$ derivatives are not covariant. 
Introducing $\ff A_k = i (\partial_k \ff U) \ff U^\dagger$ and the covariant derivative $D_k \widetilde{\ff H} = \partial_k \widetilde{\ff H} + i [ \ff A_k , \widetilde{\ff H} ]$, one instead has $D_k \widetilde{\ff H} = \ff U (\partial_k \ff H) \ff U^\dagger$. 
Hence the singular condition reads as $D_k \ff H=0$, and therefore the location of an immersion-driven transition, is representation independent. 
In the natural orbital representation used throughout this work, the connection vanishes and $D_k \ff H = \partial_k\ff H$.

Concluding, Eq.\ (\ref{eq:nusym}) explains that if $\nu = 0$ for $t_{2} < t_{1}$, we must have $\nu = 1$ (if defined) for $t_{2} > t_{1}$ and vice versa, see Fig.\ \ref{fig:pd}.
Furthermore, if $\nu$ were well-defined for $t_{1}=t_{2}$, the relation would imply $\nu = 1/2$, which is impossible since $\nu \in \mathbb{Z}$. 
Therefore, for any disorder strength $W$, where the winding number is well-defined on both sides of the line, $\nu$ must be ill-defined on the $t_{1}=t_{2}$ line, and thus this line is a topological-transition line.

\mbox{} \newline \pp{Section E: Perturbative approach to the small-$t_{2}$ regime}  \newline

\noindent
We consider the SSH model with independent binary disorder, i.e., with on-site energies $\epsilon_{i\rm A}=z_{i\rm A}W/2$ and $\epsilon_{i\rm B}=z_{i\rm B}W/2$, where $z_{iA}=\pm1$ and $z_{iB}=\pm1$ with equal probability. 
Here, we treat the intercell hopping $t_2$ perturbatively. 

For $t_2=0$, the system decomposes into independent dimers. 
At zero frequency, the Green's function of a dimer is
\be
\ff g
=
\frac{4}{z_Az_BW^2-4t_1^2}
\begin{pmatrix}
z_BW/2 & t_1\\
t_1 & z_AW/2
\end{pmatrix}
\: .
\ee
Averaging over the four configurations $z_A, z_B = \pm 1$ gives
\be
\overline {\ff g }
= 
a(W) \ff \sigma_x
\: ,
\ee
with
\be
a \equiv a(W) = \frac{16t_1^3}{W^4-16t_1^4}
\: .
\labeq{aw}
\ee

To determine the correction of first order in $t_{2}$, we consider the intercell hopping term in reciprocal space
\be
\ff V(k)
=
t_2
\left(
\cos k \, \ff \sigma_x + \sin k \, \ff \sigma_y
\right)
\: .
\ee
To first order in $t_2$, the resolvent expansion yields
\be
\overline{\ff G(k,0)}
=
\overline{\ff g} + \overline{\ff g} \, \ff V(k) \, \overline{\ff g}
+
\ca O(t_2^2)
\: ,
\ee
where the factorization of the disorder average in the second term follows from the statistical independence of the on-site energies in neighboring unit cells. Hence, with $a=a(W)$ from \refeq{aw}, 
\be
\overline{\ff G}(k,0)
=
\left[
a + a^2 t_2 \cos k 
\right] \ff \sigma_x
- a^2 t_2 \sin k \, \ff \sigma_y
+
\ca O(t_2^2)
\: .
\ee
Using the identity $(g_{x} \ff \sigma_{x} + g_{y} \ff \sigma_{y})^{-1} = (g_{x} \ff \sigma_{x} + g_{y} \ff \sigma_{y}) / (g_{x}^{2} + g_{y}^{2})$, we find the topological Hamiltonian
\be
\ff H_{\rm top}(k) 
= 
- \overline{\ff G}(k,0)^{-1}
=
d_x(k) \ff \sigma_x+d_y(k) \ff \sigma_y 
\ee
with
\be
d_x(k)
=
-
\frac
{a+a^2t_2\cos k}
{a^2+2a^3t_2\cos k+a^4t_2^2}
+
\ca O(t_2^2),
\labeq{tdx}
\ee
and
\be
d_y(k)
=
\frac
{a^2t_2\sin k}
{a^2+2a^3t_2\cos k+a^4t_2^2}
+
\ca O(t_2^2)
\: .
\labeq{tdy}
\ee

For a regular closed planar curve, the turning number is the winding number of its tangent vector and can be written as
\be
\tau
=
\frac{1}{2\pi i}
\int_0^{2\pi}dk \,
\frac
{\partial_k^2d_+(k)}
{\partial_kd_+(k)}
\: , 
\ee
where $d_+(k)=d_x(k)+id_y(k)$. 
With \refeq{tdx} and \refeq{tdy} we find the compact expression
\be
d_+(k)
=
- \frac{1}{a+a^2t_2e^{ik}} 
=
- \frac{1}{f(k)} 
\: 
\labeq{df}
\ee
with $f(k) \equiv a+a^2t_2e^{ik}$ and suppressing corrections $\ca O(\tau_{2}^{2})$. 
Now
\be
\partial_k d_+(k)
=
\frac{\partial_k f(k)}{f(k)^2}
\ee
and 
\ba
\frac
{\partial_k^2 d_+(k)}
{\partial_k d_+(k)}
&=&
\partial_{k}
\ln \left[ \partial_{k} d_{+}(k) \right]
=
\partial_{k}
\ln \left[ 
\frac{\partial_k f(k)}{f(k)^2}
\right]
\nonumber \\
&=&
\frac
{\partial_k^2f(k)}
{\partial_kf(k)}
- 2 
\frac
{\partial_k f(k)}
{f(k)}
\: .
\ea
Therefore,
\be
\tau
=
\frac{1}{2\pi i}
\int_0^{2\pi} dk \, 
\frac
{\partial_k^2f(k)}
{\partial_kf(k)}
- 
2\frac{1}{2\pi i}
\int_0^{2\pi} dk
\frac{\partial_kf(k)}{f(k)}
\: .
\labeq{tcalc}
\ee
To evaluate the first integral, we compute
\be
\frac
{\partial_k^2f(k)}
{\partial_kf(k)}
=
\frac{-a^2 t_2 e^{ik}}
{i a^2 t_2 e^{ik}}
= i
\: ,
\ee
and thus
\be
\frac{1}{2\pi i}
\int_0^{2\pi} dk \, 
\frac
{\partial_k^2f(k)}
{\partial_kf(k)}
=1
\: .
\ee
The second integral is a winding number 
\be
\nu'
=
\frac{1}{2\pi i}
\int_0^{2\pi}dk
\frac{\partial_kf(k)}{f(k)}
\: .
\ee
around the origin of the circle
\be
k \mapsto f(k) = a + a^2 t_2 e^{ik} 
\: .
\ee
With \refeq{df} we see that the winding number of the topological Hamiltonian $\nu = - \nu'$.

Combining the two terms yields
\be
\tau = 1 + 2 \nu
\: .
\ee
This circle has center $a$ and radius $|a^2t_2|$. 
Therefore,
\ba
\nu = 0  && \qquad \mbox{if} \quad |a(W)t_2|<1
\; , \nonumber
\\
\nu = -1  && \qquad \mbox{if} \quad |a(W)t_2|>1
\; , 
\ea
and consequently
\ba
\tau = +1 && \qquad \mbox{if} \quad |a(W)t_2|<1
\; , \nonumber
\\
\tau = -1 && \qquad \mbox{if} \quad |a(W)t_2|>1
\; . 
\ea
For $| a(W) t_2 | = 1$, the circle passes through the origin at $k_{\rm c} = 0$ or $k_{\rm c} = \pi$. 
This means that $\overline{\ff G}(k_{\rm c},0)$ vanishes and that $H_{\rm top}(k_{\rm c})$ diverges, so that the turning number is not defined.

For $t_1, t_2 > 0$ and $t_2 < t_1$, the two solutions of $| a(W) t_2 | = 1$ [see \refeq{aw}] are
\be
W_\pm
=
2 t_1
\left( 1 \pm \frac{t_2}{t_1} \right)^{1/4}
\: .
\ee
At $W=W_-$ one has $a t_2=-1$ and the Green-function zero occurs at $k=0$, whereas at $W=W_+$ one has $a t_2=+1$ and it occurs at $k=\pi$. 
Thus,
\ba
\nu = 0 \, , \quad \tau = +1 && \qquad \mbox{if} \quad W < W_{-}
\; , \nonumber
\\
\nu = -1 \, , \quad \tau = -1 && \qquad \mbox{if} \quad W_{-} < W < W_{+}
\; , \nonumber
\\
\nu = 0 \, , \quad \tau = +1 && \qquad \mbox{if} \quad W_{+} < W
\; . 
\ea
A change of turning number (and of the winding number) is predicted only within an interval of width $\ca O(t_2)$ around $W=2t_1$.
This narrow interval requires caution. At $W=2t_1$, the isolated-dimer Green function is singular and $a(W)$ diverges. The perturbative expansion in $t_2$ is therefore nonuniform precisely in the region where $|a(W)t_2|$ becomes of order unity. Accordingly, the values $W_\pm$ obtained from the truncated Green function cannot be regarded as quantitatively controlled phase boundaries. Nevertheless, the calculation explicitly illustrates how a change of turning number can be associated with a zero of the disorder-averaged Green function and hence with a divergence of the topological Hamiltonian.

Importantly, however, for fixed $W \neq 2 t_1$ and sufficiently small $t_2 > 0$, one obtains $\nu = 0$ and $\tau = +1$. 
This fixes the value of $\tau$ close to the $t_{2}=0$ axis in the phase diagram Fig.\ \ref{fig:pd}.
For $W \to 0$, the result is consistent with the clean-limit result for $t_{2} < t_{1}$. 
This is expected, since there is no gap closure allowing for a change of $\nu$.
Regarding the turning number, it suggests that there is also no loss of immersion allowing for a change of $\tau$.

\mbox{} \newline \pp{Section F: Perturbative approach for strong disorder}  \newline

\noindent
We consider the strong-disorder limit $W \gg t_1,t_2$. 
Let $\ff T$ with elements $T_{rr'}$ denote the hopping matrix of the clean SSH model, where $r = (i,\alpha)$ refers to the site $\alpha={\rm A, B}$ in the unit cell $i=1,...,L$. 
Furthermore, $\ff V$ with elements $V_{rr'} = \delta_{rr'} \epsilon_{r}$ is the disorder term, where $\epsilon_{r} = z_{r}W/2$ and $z_{r} = \pm 1$ with equal probabilites.
The real-space representation of the Green's function then reads
\be
  \ff G(\omega) = \frac{1}{\omega - \ff T - \ff V} \: . 
\ee
Defining the (negative) on-site Green's function at zero frequency as $g_{rr'} = - \delta_{rr'} \frac{1}{\epsilon_{r}} = - \delta_{rr'} \frac{2z_{r}}{W}$, we can expand the Green's function as
\ba
\ff G(0) 
&=& 
( \ff g^{-1} - \ff T )^{-1} 
\nonumber \\
&=& \ff g + \ff g \ff T \ff g + \ff g \ff T \ff g \ff T \ff g + \ff g \ff T \ff g \ff T \ff g \ff T \ff g + \cdots 
\: .
\labeq{locator}
\ea
Upon disorder averaging, 
\be
\overline{g_r} = 0 \: , \qquad
\overline{g_r^2} = \frac{4}{W^2} \: .
\ee
Furthermore, 
\be
\overline{g_r g_{r'}} = 0 \quad (r \ne r')
\: , 
\ee
for local and uncorrelated disorder. 
The second term in the expansion \refeq{locator} vanishes, since $\ff T$ has no diagonal entries. 
All terms with an odd power in $\ff g$ must vanish as well.
The leading nonvanishing contribution connecting two neighboring sites is thus of third order in the hopping and corresponds to traversing the same bond three times,
\be
\overline{
g_r T_{rr'} g_{r'} T_{r'r} g_{r} T_{rr'} g_r'
}
=
\frac{16}{W^4} T_{rr'}^3
\: 
\ee
with $T_{rr'}=T_{r'r}$.
After Fourier transformation, we thus have to leading order in $1/W$
\be
\overline{\ff G}(k,0)
=
\frac{16}{W^4}
\left[
\left(
t_1^3+t_2^3\cos k
\right) 
\ff \sigma_x
+
t_2^3\sin k \, 
\ff \sigma_y
\right]
+
\ca O(\frac{1}{W^{6}})
\: .
\ee
Exploiting the identity $(g_{x} \ff \sigma_{x} + g_{y} \ff \sigma_{y})^{-1} = (g_{x} \ff \sigma_{x} + g_{y} \ff \sigma_{y}) / (g_{x}^{2} + g_{y}^{2})$, we get the topological Hamiltonian
\be
\ff H_{\rm top}(k) 
= 
- \overline{\ff G}(k,0)^{-1}
=
d_x(k) \ff \sigma_x+d_y(k) \ff \sigma_y 
\: ,
\ee
where
\ba
d_{x}(k) &=& - \frac{W^{4}}{16} 
\frac
{t_{1}^{3} + t_{2}^{3} \cos k}{t_{1}^{6} + t_{2}^{6} + 2 t_{1}^{3} t_{2}^{3} \cos k}
\: , 
\nonumber \\
d_{y}(k) &=& - \frac{W^{4}}{16} 
\frac
{t_{2}^{3} \sin k}{t_{1}^{6} + t_{2}^{6} + 2 t_{1}^{3} t_{2}^{3} \cos k}
\: .
\ea
We note that apart from the overall factor $W^{4}/16$ the Bloch vector $\ff d(k)$ becomes $W$ independent as $W \to \infty$.
Introducing 
\be
d_{+}(k) = d_{x}(k) + i d_{y}(k) 
\: , 
\ee
we find the simple expression
\be
d_{+}(k) 
=
- \frac{W^{4}}{16} 
\frac{1}{t_{1}^{3} + t_{2}^{3} e^{-ik}} 
=
- \frac{W^{4}}{16} \frac{1}{f(k)}
\ee
with $f(k) \equiv t_{1}^{3} + t_{2}^{3} e^{-ik}$. 

The winding number is obtained via
\be
\nu
=
- \frac{1}{2\pi i}
\int_0^{2\pi} dk
\frac{\partial_k f(k)}{f(k)}
=
\frac{1}{2\pi i}
\oint_{|z|=1}
\frac{dz}
{z+t_1^3/t_2^3}
\: ,
\ee
with $z=e^{-ik}$ and where the contour is mathematically positively oriented.
With the residue theorem, we get $\nu = +1$
if the pole $z_{0} = - t_{1}^{3}/t_{2}^{3}$ lies inside the circle and $\nu = 0$ else: 
\ba
\nu = +1  && \qquad \mbox{if} \quad t_{2} > t_{1}
\; , \nonumber
\\
\nu = 0  && \qquad \mbox{if} \quad t_{2} < t_{1}
\; ,
\ea
where $t_{1}, t_{2} >0$.
This is the same result as in the clean limit. 

The turning number can be calculated via
\be
\tau
=
\frac{1}{2\pi i}
\int_0^{2\pi} dk \, 
\frac
{\partial_k^2f(k)}
{\partial_kf(k)}
- 
2\frac{1}{2\pi i}
\int_0^{2\pi} dk
\frac{\partial_kf(k)}{f(k)}
\, ,
\ee
see \refeq{tcalc}.
We have
\be
\frac{1}{2\pi i}
\int_0^{2\pi} dk \, 
\frac
{\partial_k^2f(k)}
{\partial_kf(k)}
=
\frac{1}{2\pi i}
\int_0^{2\pi} dk \, 
(-i)
=
- 1
\: .
\ee
Hence, $\tau = - 1 + 2 \nu$, and thus
\ba
\tau = +1  && \qquad \mbox{if} \quad t_{2} > t_{1}
\; , \nonumber
\\
\tau = -1  && \qquad \mbox{if} \quad t_{2} < t_{1}
\: 
\ea
(for $t_{1}, t_{2} >0$) in the strong-disorder limit.

For $t_2\ll t_1$, this result appears at first sight to contradict the expansion in $t_2$ discussed above. 
In that expansion, the topological-Hamiltonian curve is described by
\be
d_+(k)
=
- \frac{1}{a+a^2t_2e^{ik}}
\: ,
\ee
where
\be
a = a(W) = \frac{16t_1^3}{W^4-16t_1^4}
\: .
\labeq{aaw}
\ee
Expanding this expression to first order in $t_2$ gives
\be
d_+(k)
=
- \frac{1}{a}
+ t_2 e^{ik}
+ \ca O (t_2^2)
\: .
\ee
Thus, the leading $k$-dependent contribution has positive orientation and yields $\tau=+1$. 
With $W \gg t_1$ and \refeq{aaw}, we have
\be
d_+(k)
\simeq
- \frac{W^4}{16t_1^3}
+ t_2 e^{ik}
\: .
\ee

On the other hand, the strong-disorder expansion gives
\be
d_+(k)
=
-\frac{W^4}{16}
\frac{1}{t_1^3+t_2^3 e^{-ik}}
+ \cdots 
\: .
\ee
For $t_2<t_1$, expansion in $t_2/t_1$ yields
\be
d_+(k)
=
-\frac{W^4}{16 t_1^3}
+
\frac{W^4 t_2^3}{16t_1^6} e^{-ik}
+
\ca O \left(
\frac{W^4t_2^6}{t_1^9}
\right)
\: .
\ee
The leading $k$-dependent contribution therefore has negative orientation and yields $\tau=-1$. 
Although this term is of third order in $t_2$, its coefficient grows as $W^4$, whereas the term linear in $t_2$ obtained from the small-$t_2$ expansion is independent of $W$.

Retaining the leading $k$-dependent contribution from both expansions, we may therefore write asymptotically
\be
d_+(k)
\simeq
d_0
+
B e^{ik}
+
C e^{-ik}
\: ,
\ee
with
\be
d_0 \simeq
- \frac{W^4}{16t_1^3}
\, , 
\qquad
B = t_2 \, ,
\qquad
C = \frac{W^4t_2^3}{16t_1^6}
\: .
\ee
The tangent vector is then
\be
\partial_k d_+(k)
\simeq
iB e^{ik}
-iC e^{-ik}.
\ee
For $B>C$, the positively oriented contribution dominates and the tangent winds once counterclockwise, i.e., 
$\tau=+1$.
For $C>B$, the negatively oriented contribution dominates, i.e., $\tau=-1$.
The change of turning number occurs when $B=C$, namely at the critical intercell hopping
\be
t_{2,\rm c}
\simeq
\frac{4 t_1^3}{W^2}
\: ,
\qquad
W \gg t_1
\: .
\ee
Writing the condition in the form
\be
\frac{t_{2,\rm c}}{t_1}
\simeq
4\left(\frac{t_1}{W}\right)^2
\ll
1
\: , 
\ee
shows that the critical line indeed lies within the small-$t_2$ regime for sufficiently strong disorder.

Right at the transition, $B=C$, we have
\be
\partial_k d_+(k)
=
iB\left(e^{ik}-e^{-ik}\right)
=
-2B\sin k
\: . 
\ee
which vanishes at $k=0$ and $k=\pi$. 
The topological-Hamiltonian curve thus ceases to be an immersion at the transition. In contrast, $d_+(k)$ itself remains finite in the regime $t_{2,\rm c}\ll t_1$, since the large constant term $d_0$ dominates over both oscillating contributions. Within this asymptotic analysis, the change from $\tau=+1$ to $\tau=-1$ is therefore associated with a vanishing tangent rather than with a divergence of the topological Hamiltonian.

\mbox{} \newline \pp{Section G: Consequences of time-reversal symmetry}  \newline

\noindent
For a given disorder realization $\hat V$, the single-particle Hamiltonian $\hat{H} = \hat{H}_{0} + \hat V$ is time-reversal invariant, 
$[\hat T, \hat H] = 0$, where, for spinless particles, $\hat T$ is the antiunitary representation given by complex conjugation in the real-space basis.
Hence, for the clean-limit Bloch Hamiltonian in reciprocal space, time-reversal invariance enforces
\be
\ff H_{0}(k)^{\ast} = \ff H_{0}(-k)
\: .
\ee
Since the disorder probability distribution is invariant under time reversal as well, we find for the disorder-averaged Green's function
\be
\ff G(k,\omega)^{\ast} = \ff G(-k,\omega^{\ast}) 
\: . 
\ee
For the topological Hamiltonian $\ff H_{\rm top}(k) = - \ff G^{-1}(k,0)$, this implies
\be
\ff H_{\rm top}(k)^\ast = \ff H_{\rm top}(-k) 
\: ,
\labeq{shtop}
\ee
as long as the Green's function is nonsingular and has no spectral weight at $\omega=0$. 

The topological Hamiltonian can be decomposed as
\be
\ff H_{\rm top}(k)
= d_x(k) \ff \sigma_x
+ d_y(k) \ff \sigma_y
\: ,
\ee
where the $k$-dependent coefficients $d_x(k)$ and $d_y(k)$ are real.
Note that chiral symmetry (see Sec.\ D) enforces $d_{0}(k) \equiv 0$ and $d_{z}(k) \equiv 0$. 
Using $\ff \sigma_x^\ast = \ff \sigma_x$ and $\ff \sigma_y^\ast = - \ff \sigma_y$, time-reversal symmetry, i.e., \refeq{shtop}, gives us 
\ba
d_x(k) &=&  d_x(-k) \: , 
\nonumber \\
d_y(k) &=& -d_y(-k) \: .
\labeq{condd}
\ea
The closed curve ${\ff d}(k) = (d_x(k) , d_y(k))$ is therefore symmetric under reflection at the $d_x$ axis. 
In particular,
\be
d_y(0) = d_y(\pi) = 0 
\: .
\ee
Let $\ff v(k) = \partial_{k} \ff d(k)$ denote the tangent at $k$.
Differentiation of \refeq{condd} with respect to momentum yields
\ba
v_x(k) &=& - v_x(-k) \: , 
\nonumber \\
v_y(k) &=& v_y(-k) \: 
\labeq{condv}
\ea
Hence,
\be
v_x(0)=v_x(\pi)=0 
\: .
\labeq{v0}
\ee

The turning number of the planar curve is the winding number of its tangent vector,
\be
\tau 
=
\frac{1}{2\pi} \int_{\rm BZ} dk \, 
\partial_k \arg
\left[
\partial_k d_x(k) + i \partial_k d_y(k)
\right] 
\: .
\ee
It is invariant under regular homotopies and can change only if the curve ceases to be an immersion, i.e., if for some critical momentum and disorder strength
\be
\partial_k{\bf d}(k_{\rm c},W_{\rm c})=0 
\: .
\ee
The consequences of time-reversal symmetry depend qualitatively on whether $k_{\rm c}$ is a time-reversal-invariant momentum.

We first consider a generic critical momentum $k_{\rm c} \neq 0 , \pi$. 
Equation (\ref{eq:condv}) implies that, if $\ff v(k_{\rm c},W_{\rm c}) = 0$ then necessarily $\ff v(-k_{\rm c},W_{\rm c}) = 0$.
A generic immersion singularity at $k_{\rm c} \neq 0 , \pi$ is therefore accompanied by a second singularity at the mirror-related momentum $-k_{\rm c}$, and both occur at the same disorder strength $W_{\rm c}$.
One can show that both singularities contribute equally to the change of the turning number.
To this end we define the complex tangent as 
\be
u(k,W) = v_x (k,W) + i v_y(k,W)
\: .
\ee
Using Eqs.\ (\ref{eq:condv}), we get 
\ba
u(k,W)^\ast  
&=&
v_x (k,W) - i v_y(k,W)
\nonumber \\
&=&
- v_x (-k,W) - i v_y(-k,W)
= 
-u(-k,W)
\: .
\nonumber \\
\labeq{com}
\ea
Note that $u(k,W)^\ast$ is the mirror of the tangent $u(k,W)$ and that, following \refeq{com}, complex conjugation reverses the orientation in the complex tangent plane, while the transformation $k \to -k$ reverses the orientation in momentum space. 
The two orientation reversals cancel, explaining why the two mirror-related singularities carry the same local index.
Consequently, we must have $\Delta \tau = \pm 2$.
Thus, time-reversal symmetry naturally permits a direct transition between $\tau=+1$ and $\tau=-1$ at a single critical disorder strength, without an intervening phase with $\tau=0$. 

The situation is different when the immersion is lost at a time-reversal-invariant momentum, $k_{\rm c}=0$ or $k_{\rm c} = \pi$. 
Such a momentum is its own mirror partner, so that no second singularity is generated by symmetry. 
Moreover, $\partial_k d_x(k_{\rm c})=0$ is already enforced by time-reversal symmetry, see \refeq{v0}, and the loss of immersion requires only
\be
\partial_k d_y(k_{\rm c},W_{\rm c}) = 0 
\: .
\ee
A generic singularity at $k_{\rm c}=0$ or $\pi$ therefore changes the turning number by one unit,
\be
\Delta \tau = \pm 1 
\: .
\ee
Consequently, a transition from $\tau=+1$ to $\tau=-1$ mediated exclusively by singularities at time-reversal-invariant momenta generically requires two distinct critical disorder strengths,
\be
\tau=+1
\:
\stackrel{W_{{\rm c}1} \quad \!\!\!}{\longrightarrow}
\:
\tau=0
\:
\stackrel{W_{{\rm c}2} \quad \!\!\!}{\longrightarrow}
\:
\tau
= - 1 
\: .
\ee

Thus, while both types of immersion singularities are locally generic, time-reversal symmetry distinguishes two different scenarios for a total change $\Delta\tau=\pm2$: either two separate transitions with an intermediate $\tau=0$ phase when the singularities occur at $k=0$ or $\pi$, or a single direct transition generated by a simultaneous mirror-related pair at $\pm k_{\rm c}$ with $k_{\rm c}\neq 0,\pi$.

\end{document}